\documentstyle[aps,prb]{revtex}
\begin{document}
\draft
%@twocolumnfalse\endcsname

%\preprint{Preprint }
\title{Fermion zero modes on vortices in  chiral superconductors. }
\author{G.E. Volovik\\
Helsinki University of Technology,
Low Temperature Laboratory,\\
P.O.Box 2200,
FIN-02015 HUT,  Finland, \\
 Landau Institute for Theoretical Physics,
117334 Moscow,
Russia }

\date{\today}

\maketitle

\begin{abstract}
The energy levels of the fermions bound to the vortex core are
considered for the general case of chiral superconductors. There are
two
classes of chiral superconductivity: in the superconducting state of
class I
the axisymmetric singly quantized vortex  has the same energy
spectrum of
bound states as in
$s$-wave superconductor:
$E=(n+1/2)\omega_0$ with integral $n$. In the class II the
corresponding
spectrum is $E=n\omega_0$ and thus contains the state with exactly
zero energy.  The effect
of a single impurity on the spectrum of bound state is also
considered.
For the class I the spectrum  acquires the double period
$\Delta E=2\omega_0$ and consists of two equidistant sets of levels in
accordance with   Larkin and
Ovchinnikov in  Phys. Rev. {\bf B57},  5457 (1998)
\cite{LarkinOvchinnikov}.
The spectrum is not influenced by a single impurity if the same
approximation
is applied for the class II states.

\end{abstract}
\


%\eject
%]
%\narrowtext
%\twocolumn

{\bf Introduction.}
The low-energy fermions bound to the vortex core play the main role
in the
thermodynamics and dynamics of the vortex state in superconductors and
Fermi-superfluids. The spectrum of the low-energy bound states in the
core
of the
axisymmetric vortex with winding number $m=\pm 1$ in the isotropic
model of
$s$-wave superconductor was obtained in microscopic theory by Caroli,
de
Gennes and Matricon
\cite{Caroli}:
\begin{equation}
E_n=\omega_0\left(n+{1\over 2}\right).
 \label{Caroli}
\end{equation}
This spectrum is two-fold degenerate due to spin degrees of freedom.
The
integral quantum number $n$ is related to the  angular
momentum of the fermions $n=-m L_z$.
The level spacing is small compared to the energy gap of the
quasiparticles
outside the core, $\omega_0\sim \Delta^2/E_F\ll\Delta$. So, in many
physical
cases the discreteness of $n$ can be neglected and one can apply the
quasiclassical approach to calculate the energy spectrum. Using this
simplified approach one obtaines that the spectrum crosses zero
energy as a
function of continuous angular momentum $L_z$. So, one has the
fermion zero
modes. The fermions in this 1D "Fermi liquid" are chiral: the positive
energy fermions have a definite sign of the angular momentum $L_z$. In
general case of arbitrary winding number $m$, the number of fermion
zero
modes, i.e. the number of branches crossing zero level, equals $-2m$
(see Ref.\cite{Q-modes-Index}). This represents an analogue of the
index
theorem known in the relativistic quantum field theory.

At low temperature $T\sim\omega_0$ the discrete nature of the spectrum
becomes important. The quantization of the zero modes was obtained
within the quasiclassical approach using the Bohr\,--\,Sommerfeld
scheme \cite{KopninVolovik}. However the term ${1/2}$ in
Eq.(\ref{Caroli}), which came from the phase shift, was missing in
this
approach and was restored only after using the general symmetry
arguments in
Ref.\cite{MissirpashaevVolovik}. Here we extend the quasiclassical
approach and obtain an exact quantization rule. We find that with
respect
to the phase shift the superconducting/superfluid states can be
divided into
2 classes:  In the states of class I the spectrum of bound
states in the
$m=\pm 1$ vortices is the same as in
Eq.(\ref{Caroli}). In the states of class II the
corresponding spectrum is:
\begin{equation}
E_n=\omega_0 n ~,
 \label{CaroliII}
\end{equation}
it contains the state with $n=0$, which has exactly
zero energy. The repesentatives of this class are the
superfluid
$^3$He-A, where the existence of the zero-energy bound state was
first calculated by Kopnin and Salomaa in a microscopic
theory \cite{KopninSalomaa}, and possibly the layered
superconductor Sr$_2$RuO$_4$, where the chiral
$p$-wave superconductivity is suggested~\cite{Rice}.

Using the quasiclassical
approach we also consider how the spectrum changes under the
influence of a
single impurity in the vicinity of the vortex core. We find that the
Larkin\,--\,Ovchinnikov result obtained for $s$-wave vortex
\cite{LarkinOvchinnikov,KoulakovLarkin2} is valid only for the class I
of superconducting states.

{\bf Quasiclassical approach to bound states in the vortex core.}
 In this
approach developed in
\cite{Q-modes-Index,Stone,KopninVolovik},  the fast radial motion of
the
fermions in the vortex core is integrated out and one obtains only
the slow
motion corresponding to the fermion zero modes. Since many properties
of
the fermion zero modes do not depend on the exact structure of the
order
parameter and vortex core,  we consider for simplicity the following
pairing states
\begin{eqnarray}
{\rm spin~singlet}:~~{\bf \Delta}=\Delta({\bf r})(\hat p_x+i\hat
p_y)^N\hat p_z^{l-|N|} ~, ~{\rm odd}~l~,
\label{OrderParameterSinglet}
\end{eqnarray}
\begin{eqnarray}
{\rm spin~triplet}:~~{\bf \Delta}=\sigma_z\Delta({\bf r})(\hat
p_x+i\hat
p_y)^N\hat p_z^{l-|N|} ~,~{\rm even}~l.
 \label{OrderParameterTriplet}
\end{eqnarray}
 Here ${\hat {\bf p}}$ is the direction of the quasiparticle
momentum; $N$ and $l\geq |N|$ are integers; $\sigma_z$ is the
Pauli matrix for conventional spin. The chiral superconductors are
characterized by the nonzero value of the index
$N$, which in our simple case represents the projection of the orbital
angular momentum
$l$ of Cooper pair along the axis $z$. For example,  the
$s$-wave superconductor has numbers $N=l=0$ in
Eq.(\ref{OrderParameterSinglet}), while the triplet $p$-wave
($l=1$) superconductor with the order parameter of the $^3$He-A type
is
specified by the numbers
$|N|=l=1$ in Eq.(\ref{OrderParameterTriplet}). The index $N$ is also
the
topological invariant in the momentum space, which is responsible for
the
Chern\,--\,Simons terms in the 2D superfluids/superconductors
(see Refs.[10\,--\,12], that is why it is well
determined even  in the case, when the  momentum $l$ and its
projection
are not determined.

 We assume the following structure of the
order parameter in the core:
\begin{equation}
\Delta({\bf r}) =\Delta(r) e^{im\phi}~,
\label{OrderParameterInVortex}
\end{equation}
where $z$, $r$, $\phi$ are the coordinates of the cylindrical system
with
the axis $z$ along the vortex line.

The Bogoliubov\,--\,Nambu Hamiltonian for quasiparticles
is given by
\begin{equation}
{\cal H}_{\bf p}=\left(\matrix{{p^2-p_F^2\over 2m_e} & {\bf \Delta}\cr
{\bf \Delta}^*&-{p^2-p_F^2\over 2m_e}\cr}\right)~.
 \label{MicroHamiltonian}
\end{equation}
In the quasiclassical approach it is assumed that the characteristic
size
$\xi$ of the vortex core is much larger than the wave length: $\xi
p_F\gg
1$. In this quasiclassical limit the description in terms of
trajectories is
most relevant. The trajectories are almost the straight lines. The low
energy trajectories are characterized by the momentum ${\bf q}$ of the
incident quasiparticle on the Fermi surface, i.e.~with  $|{\bf
q}|=p_F$, and
the impact parameter $b$. Let us consider for simplicity the 2D or
layered
superconductors, so that ${\bf q}=p_F(\hat {\bf x}\cos\theta +$ $+
\hat {\bf
y}\sin\theta)$. Then substituting $\Psi\rightarrow e^{i{\bf
q}\cdot{\bf
r}}\Psi$ and ${\bf p}\rightarrow {\bf q}-i{\bf \nabla}$, and
expanding in
small ${\bf \nabla}$, one obtains the quasiclassical Hamiltonian for
the
fixed trajectory ${\bf q},b$:
\begin{equation}
{\cal H}=-i\tau_3{\bf v}_F\cdot{\bf \nabla}  +
 \Delta(r)\left(\tau_1     \cos(N\theta + m\phi)  -\tau_2
\sin(N\theta  + m\phi)\right)~,~{\bf v}_F={{\bf q}\over m_e}~.
 \label{QuasiclassicalHamiltonian}
\end{equation}
We omitted spin indices, since they are not important for the
spectrum in superconducting states under consideration.

Since the spatial derivative is along the trajectory
it is convinient to choose the
coordinate system: $s=r\cos (\phi-\theta)$ -- the coordinate
along the trajectory, and
$b=r\sin (\phi-\theta)$ (see e.g. \cite{KopninSalomaa}). In this
system the Hamiltonian is
\begin{eqnarray}
%\begin{equation}
 {\cal H} = -i{v}_F\tau_3 \partial_s +
 \tau_1  \Delta(r)   \cos \left(m \tilde\phi +(m+N)\theta\right)-
\nonumber\\
-\tau_2    \Delta(r)
  \sin \left(m \tilde\phi +(m+N)\theta \right)  ~,~~~\tilde\phi=\phi
-\theta~.
 \label{QuasiclassicalHamiltonian2}
\end{eqnarray}
%\end{equation}
The dependence of the Hamiltonian on
the direction $\theta$ of the trajectory can be removed by the
following
transformation:
\begin{eqnarray}
\Psi=
e^{i (m+N)\tau_3 \theta/2}\tilde\Psi,
\label{TransformationFunction}
\end{eqnarray}
$$\tilde{\cal H}=e^{-i (m+N)\tau_3 \theta/2}{\cal H}e^{i (m+N)\tau_3
\theta/2}=
  -i{v}_F\tau_3 \partial_s +$$
\begin{eqnarray}
+\Delta(\sqrt{s^2+b^2})\left(\tau_1  \cos  m
\tilde\phi  -\tau_2    \sin  m \tilde\phi\right),
\end{eqnarray}
\begin{eqnarray}
\tan
\tilde\phi={b\over s} .
 \label{TransformationHamiltonian}
\end{eqnarray}
Now $\theta$ enters only the boundary
condition for the wave function, which according to
Eq.(\ref{TransformationFunction}) is
\begin{equation}
\tilde\Psi(\theta +2\pi)=(-1)^{m+N}\tilde\Psi(\theta)
 \label{BoundaryCondition}
\end{equation}
With respect to this boundary condition, there are two classes of
vortices: with odd and even $m+N$. The
quantum spectrum of fermions in the core is essentially
determined by this condition. Let us consider vortices
with
$m=\pm 1$.

The quasiclassical Hamiltonian in
Eq.(\ref{TransformationHamiltonian}) is
the same as for the $s$-wave vortex and thus can be treated in the
same
manner as in Ref.\cite{Q-modes-Index}. The state with the
lowest energy corresponds to trajectories, which cross the center of
the
vortex, i.e. with $b=0$. Along this trajectory one has
$
\sin
\tilde\phi=0$ and     $\cos   \tilde\phi = {\rm sign}~ s$. So that
the Eq.(\ref{TransformationHamiltonian})
\begin{equation}
 \tilde{\cal H} = -i{v}_F\tau_3 \partial_s +
 \tau_1  \Delta(|s|)  {\rm sign}~ s
 \label{SupersymmetriclHamiltonian}
\end{equation}
becomes supersymmetric and thus contains the eigenstate with zero
energy. Let us write the corresponding eigen function  including all
the transformations:
\begin{equation}
 \Psi_0(s,\theta, b=0) =  e^{ip_F s} e^{i (m+N)\tau_3
\theta/2} \left(\matrix{1 \cr -i\cr}\right)
\psi_0(s)~~,~~\psi_0(s)=\exp{\left(-\int^s ds' {\rm sign}~
s'{\Delta(|s'|)\over v_F}
\right)}~.
 \label{SupersymmetriclSolution}
\end{equation}
When $b$ is small the third term in
Eq.(\ref{TransformationHamiltonian}) can
be considered as perturbation and this gives the energy levels in
terms of
$b$ and thus in terms of the angular momentum $L_z=p_Fb$:
\begin{equation}
E(L_z,\theta)=-mL_z \omega_0~,~\omega_0  = {\int_{-\infty}^\infty ds
|\psi_0(s)|^2 {
\Delta(|s|)\over p_F|s|}\over \int_{-\infty}^\infty ds
|\psi_0(s)|^2}
 \label{QuasiclasicalEnergy}
\end{equation}

The next step is the quantization of motion in the $\theta,L_z$ plane.
Since the angle and momentum are canonically conjugated variable, the
quantized energy levels are obtained from the quasiclassical
energy in Eq.(\ref{QuasiclasicalEnergy}), if $L_z$ is considered as an
operator. The Hamiltonian
\begin{equation}
H(\theta)= im\omega_0\partial_\theta
 \label{HamiltonianTheta}
\end{equation}
has the eigenfunctions $e^{-iE\theta/m\omega_0}$.
The boundary condition, the
Eq.(\ref{BoundaryCondition}), gives the following quantization of the
energy levels for $m=\pm 1$ vortices
\begin{equation}
E_n =  n \omega_0,~~{\rm odd}~~{N};~~
E_n =  \left(n+{1\over 2}\right) \omega_0,~~{\rm even}~~{N}.
\label{Spectrum}
\end{equation}

{\bf Effect of single impurity.}
  As distinct from the Andreev scattering in the vortex core, which
leads
to the bound states, the microscopic impurity leads to the
conventional
elastic scattering in which the momentum ${\bf q}$ of quasiparticle
changes
and thus the transition between different trajectories occurs. In the
limit
of low energy of the quasiparticle the impact parameter $b$ of the
scattered
particle is close to zero and thus is smaller than the distance $R$
from
impurity to the center of the vortex, which is of order $\xi$. If we
assume the atomic size of impurity, then the scattering of the low-
energy
quasiparticle occurs only between two trajectories which cross
simultaneously the vortex center and impurity. Thus we are interested
in
the matrix element between the states with
$\theta=\theta_{imp}$ and   $\theta=\pi + \theta_{imp}$
[8]. In the general case this coupling has a form
\begin{equation}
{\cal H}_{imp}=2\lambda e^{i\gamma}\psi(\pi+ \theta_{
imp})\psi^*(\theta_{imp}) + 2\lambda e^{-i\gamma}
\psi^*(\pi+
\theta_{imp})\psi(\theta_{imp})~.
 \label{JosephsonCoupling}
\end{equation}

Together with free Hamiltonian in Eq.(16) this gives the following
Schr\"{o}dinger equation for the motion in $\theta $-space:
$$im\omega_0{\partial \psi \over \partial\theta} +2\lambda e^{i\gamma}
\delta(\theta-\theta_{imp}) \psi(\pi+ \theta_{imp}) +$$
\begin{equation}
+2\lambda e^{-i\gamma}
\delta(\theta-\pi-\theta_{ imp})\psi(\theta_{imp}) =E\psi(\theta)~~ ,
\label{HamiltonianImpurity} \end{equation} with boundary condition
\begin{equation} \psi(\theta+2\pi) =\pm \psi (\theta)~.
\label{Boundary2}
\end{equation} Here the sign $-$ and $+$ is for the  $|m|= 1$ vortex
in class
I and class II superconducting states correspondigly.  The solution
of these
equations give the energy eigenvalues:  \begin{eqnarray} \cos {\pi
E\over
\omega_0} = {2\omega_0 \lambda\over \omega_0^2+
\lambda^2}\sin(m\gamma)
~~,~~|m|=1~~,~~{\rm class~I}~, \label{EigenValuesS} \\ \sin {\pi
E\over
\omega_0} = {2\omega_0 \lambda\over \omega_0^2+ \lambda^2}\cos\gamma
~~,~~|m|=1~~,~~{\rm class~II}~.  \label{EigenValuesP} \end{eqnarray}
For the
class I of superconducting states the Eq.(\ref{EigenValuesS}) is
similar to
Eq.(2.10) of Ref.\cite{KoulakovLarkin2} obtained for the $s$-wave
case:
the spectrum in the presence of impurity has the double period $\Delta
E=2\omega_0$ and consists of two equdistant sets of levels. These two
sets
transform to each other under symmetry transformation
$E\rightarrow -E$, which is the analogue of CPT-symmetry.

For the class I of states the Eq.(\ref{EigenValuesP}) also
gives two sets of levels with the alternating shift. But the two sets
are
not mutually symmetric with respect to $E=0$. This contradicts to  the
CPT-symmetry of the system. The only way to reconcile the
Eq.(\ref{EigenValuesP}) with this symmetry is to assume that because
of the CPT-symmetry either (i) there is no coupling between the two
trajectories; or (ii) the phase of the coupling is fixed,
$\gamma=\pi/2$. Then the energy
levels are
$E_n=n\omega_0$, i.e.  the same as without impurities. Thus the same
CPT-symmetry, which is responsible for the eigenstate with $E=0$,
provides the rigidity of the spectrum.

Now we shall show that in the
simplest model of the impurity potential $H_{imp}=$\linebreak
$=U\tau_3
\delta({\bf r}-{\bf R})$ the coupling between the opposite
trajectories does
disappear for the superconducting states of class II. Let us
consider the lowest energy trajectories; they cross the center of the
vortex, as a result
$\delta({\bf
r}-{\bf R})=\delta(s-R)\delta(\theta-\theta_{imp})/R$. The matrix
element between two wave functions in
Eq.(\ref{SupersymmetriclSolution})
corresponding to the opposite trajectories, i.e. which angles $\theta$
differ by
$\pi$, is proportional to
\begin{equation}
\lambda e^{i\gamma }\sim  {U\over R\xi}e^{2ip_F R} \exp{\left(-{2\over
v_F}\int^R_0 dr
\Delta(r)
\right)} \left(\matrix{1 &i\cr}\right)
\left[\tau_3 e^{i (m+N)\tau_3
\pi/2} \right]\left(\matrix{1 \cr -i\cr}\right) ~.
 \label{MatrixElement}
\end{equation}
From the spinor structure it follows that the impurity scattering
between
the opposite trajectories disappears,
$\lambda=0$, for vortices with even
$m+N$. Thus in the superconducting
states of class II the spectrum of fermions in the $m=\pm 1$
vortices is not influenced by the single impurity.

{\bf Conclusion.}
The phase $(m+N)\tau_3 \theta/2$ in
Eq.(\ref{TransformationFunction}) plays the part of Berry phase. It
shows
how the wave function of quasiparticle changes, when the trajectory is
rotating by angle $\theta$. This Berry phase is instrumental for the
Bohr\,--\,Sommerfeld quantization of the energy levels in the vortex
core. It
chooses between the two possible quantizations consistent with the
CPT-symmetry of states in superconductors:   $E_n=n\omega_0$ and
$E_n=(n+1/2)\omega_0$.

The same phase is also important for the effect of single impurity on
the
spectrum of bound states. We found that if in a pure superconductor
the
spectrum is
$E_n=(n+1/2)\omega_0$, the impurity splits it into two series
according to the
Larkin\,--\,Ovchinnikov prescription
\cite{LarkinOvchinnikov,KoulakovLarkin2}. However if
the initial spectrum is
$E_n=n\omega_0$ (an example is $m=\pm 1$ vortex in the chiral
$p$-wave superconductor, where $N=1$) then the impurity does not
change this
spectrum. This rigidity of the spectrum must be taken
into account when the effect of randomness due to many impurities is
considered and new  level statistics for the fermionic spectrum in
the core
is introduced
\cite{FeigelmanSkvortsov}. The existence of the level with exactly
zero
energy can essentially change the estimation \cite{Goryo} of the
fractional
charge carried by the vortex core in chiral superconductors.

I am indebted to P. Wiegmann whose remark on the Berry
phase was extremely fruitful and to M. Feigel'man and N. Kopnin for
numerous discussions.

\end{document}